\documentclass[aps,prl,superscriptaddress,nofootinbib,twocolumn]{revtex4-1}
\usepackage{amssymb,graphicx,amsmath,array,verbatim,setspace}

\newcommand{\beq}{\begin{eqnarray}}
\newcommand{\eeq}{\end{eqnarray}}
\newcommand{\p}{\partial}

\newcommand{\bpm}{\begin{pmatrix}}
\newcommand{\epm}{\end{pmatrix}}
\newcommand{\Z}{\mathbb{Z}}
\newcommand{\R}{\mathbb{R}}
\newcommand{\C}{\mathbb{C}}

\newcommand{\ba}{\left(\begin{array}}
\newcommand{\ea}{\end{array} \right)}

\usepackage[usenames]{color}

\begin{document}

\title{Confinement-Deconfinement Crossover in the Lattice $\C P^{N-1}$ Model}

\author{Toshiaki Fujimori}
\email{toshiaki.fujimori018(at)gmail.com}
\address{Department of Physics, and Research and 
Education Center for Natural Sciences, 
Keio University, 4-1-1 Hiyoshi, Yokohama, Kanagawa 223-8521, Japan}

\author{Etsuko Itou}
\email{itou(at)yukawa.kyoto-u.ac.jp}
\address{Department of Physics, and Research and 
Education Center for Natural Sciences, 
Keio University, 4-1-1 Hiyoshi, Yokohama, Kanagawa 223-8521, Japan}
\address{Department of Mathematics and Physics, Kochi University, Kochi 780-8520, Japan}
\address{Research Center for Nuclear Physics (RCNP), Osaka University, Osaka 567-0047, Japan}

\author{Tatsuhiro Misumi}
\email{misumi(at)phys.akita-u.ac.jp}
\address{Department of Mathematical Science, Akita 
University,  Akita 010-8502, Japan
}
\address{Department of Physics, and Research and 
Education Center for Natural Sciences, 
Keio University, 4-1-1 Hiyoshi, Yokohama, Kanagawa 223-8521, Japan}
\address{iTHEMS, RIKEN,
2-1 Hirasawa, Wako, Saitama 351-0198, Japan
}

\author{\\Muneto Nitta}
\email{nitta(at)phys-h.keio.ac.jp}
\address{Department of Physics, and Research and 
Education Center for Natural Sciences, 
Keio University, 4-1-1 Hiyoshi, Yokohama, Kanagawa 223-8521, Japan}

\author{Norisuke Sakai}
\email{norisuke.sakai(at)gmail.com}
\address{Department of Physics, and Research and 
Education Center for Natural Sciences, 
Keio University, 4-1-1 Hiyoshi, Yokohama, Kanagawa 223-8521, Japan}
\address{iTHEMS, RIKEN,
2-1 Hirasawa, Wako, Saitama 351-0198, Japan
}

\begin{abstract}
The $\C P^{N-1}$ sigma model at finite temperature 
is studied using lattice Monte Carlo simulations 
on $S_{s}^{1} \times S_{\tau}^{1}$ 
with radii  $L_{s}$ and $L_{\tau}$, respectively, where
the ratio of the circumferences is taken to be sufficiently large 
($L_{s}/L_{\tau} \gg 1$) to simulate the model 
on $\mathbb R \times S^1$. 
We show that the expectation value of the Polyakov loop 
undergoes a deconfinement crossover as $L_{\tau}$ is decreased, 
where the peak of the associated susceptibility 
gets sharper for larger $N$.
We find that the global PSU($N$)=SU($N$)$/{\mathbb Z}_{N}$
symmetry remains unbroken at ``quantum" and ``classical" levels  
for the small and large $L_{\tau}$, respectively:
in the small $L_\tau$ region for finite $N$, the order parameter 
fluctuates extensively with its expectation value consistent 
with zero after taking an ensemble average, while  
in the large $L_\tau$ region the order parameter 
remains small with little fluctuations.
We also calculate the thermal entropy and find that 
the degrees of freedom in the small $L_{\tau}$ regime are 
consistent with $N-1$ free complex scalar fields, 
thereby indicating a good agreement with 
the prediction from the large-$N$ study for small $L_{\tau}$.
\end{abstract}

\maketitle

{\bf Introduction}:
The ${\mathbb C}P^{N-1}$ sigma model \cite{Eichenherr:1978qa,Witten:1978bc,DAdda:1978vbw,DAdda:1978dle}
is known to show up in various aspects of physics. 
Originally, the ${\mathbb C}P^{N-1}$ model in two dimensions 
is regarded as a toy model of QCD, 
since they share various common properties 
such as asymptotic freedom, 
confinement and generation of a mass gap. 
Recently, connections between two dimensional  ${\mathbb C}P^{N-1}$ 
models and four dimensional gauge theories have been established: 
it appears as the low-energy effective theory on 
a non-Abelian vortex in the non-Abelian gauge-Higgs model 
\cite{Hanany:2003hp,Auzzi:2003fs,Eto:2005yh,Eto:2006cx,Eto:2006pg,Tong:2005un,Eto:2006pg,Shifman:2007ce}
as well as dense QCD 
\cite{Nakano:2007dr,Eto:2009bh,Eto:2009tr,Eto:2013hoa},  
on a long string of Yang-Mills theory \cite{Aharony:2013ipa}, 
and of an appropriately compactified Yang-Mills theory \cite{Yamazaki:2017ulc}. 
In condensed-matter physics, 
the ${\mathbb C}P^1$  model plays an essential role in the research 
on the low-energy behavior of anti-ferromagnetic spin chains 
and their extensions \cite{Haldane:1982rj}, 
and 
describes quantum phase transition known as 
deconfined criticality \cite{Senthil:2004aza,Nogueira:2013oza},
while  
the ${\mathbb C}P^{N-1}$ model appears as 
an $SU(N)$ spin chain \cite{Beard:2004jr}
and also can be realized in ultracold atomic gases \cite{Laflamme:2015wma,Kataoka:2010sh}.

Theoretically, non-perturbative properties
of the ${\mathbb C}P^{N-1}$ model 
have been studied both analytically 
by the gap equations 
with the large-$N$ (mean field) approximation
\cite{Witten:1978bc,DAdda:1978vbw,DAdda:1978dle}
and by lattice simulations mainly on topological aspects of the model defined on $\mathbb{R}^2$ 
\cite{Berg:1981er,Campostrini:1992ar,Alles:2000sc,Flynn:2015uma,Bruckmann:2015sua,Bruckmann:2016txt,Abe:2018loi,Bruckmann:2018rra,Bonanno:2018xtd}.
These analyses are consistent with the Coleman-Mermin-Wagner (CMW) theorem \cite{Coleman:1973ci,Mermin:1966fe} forbidding spontaneous breaking of a continuous symmetry in two dimensions, 
while perturbative analyses are not.
Recently, the large-$N$ analyses have 
been extensively applied to the  ${\mathbb C}P^{N-1}$ model 
at finite temperature, or equivalently 
the model defined on 
 $\mathbb{R} \times S^1$ 
with the periodic boundary conditions (pbc) 
 \cite{Monin:2015xwa,Monin:2016vah} 
 (see also the earlier works \cite{Hong:1994uv,Hong:1994te}).
However, these studies do not reach a consensus 
 for physics at high temperature (or at small compactification radius) 
 \cite{Monin:2015xwa,Monin:2016vah,Bolognesi:2019rwq,Flachi:2019jus} 
 (see also \cite{Nitta:2017uog,Nitta:2018lnn,Nitta:2018yen,Yoshii:2019yln}) 
 including an analogous case of the model defined on
 a finite interval 
 \cite{Milekhin:2012ca,Bolognesi:2016zjp,Milekhin:2016fai,Betti:2017zcm,Flachi:2017xat,Bolognesi:2018njt,Chernodub:2019nct,Pavshinkin:2019bed,Flachi:2019jus},
 while all studies agree that the physics at low temperature 
 (or at large radius) 
 recovers the CMW theorem.
The questions can be summarized as follows:
(i) How the order parameter is defined and
how its expectation value depends 
on the compactification period $L_{\tau}$.
(ii) How the global PSU($N$)$=$SU($N$)$/{\mathbb Z}_{N}$ symmetry is 
realized for finite $N$.
One may naively expect the global symmetry to be broken in the deconfinement phase, where field variables are ordered.
It was suggested that the global symmetry is
broken in the ``deconfinement" phase in the large-$N$ limit \cite{Monin:2015xwa,Monin:2016vah}. 
On the other hand, the CMW theorem forbids the continuous symmetry breaking at least in finite $N$.
(iii) How the high temperature behavior changes for finite $N$.
In the large-$N$ limit, an explicit high temperature behavior 
of the free energy was calculated \cite{Monin:2015xwa}.

In this Letter, we investigate the ${\mathbb C}P^{N-1}$ model 
at finite temperature by the lattice Monte Carlo simulation to solve the above mentioned questions.
We also calculate the Polyakov loop expectation value, its susceptibility and the thermal entropy. 
Our results can be summarized as follows:
(1) We adopt the absolute value of the expectation value of the Polyakov loop as a confinement-deconfinement order parameter.
We find that its $L_\tau$ dependence exhibits a crossover behavior
and the peak of its susceptibility gets sharper with $N$ increases,
implying a possibility of the phase transition in the large-$N$ limit \cite{Monin:2015xwa}. 
(2) We find that the global PSU($N$)=SU($N$)$/{\mathbb Z}_{N}$ 
symmetry remains unbroken at ``quantum" and ``classical" levels  
for the small and large $L_{\tau}$, respectively. 
(3) We calculate the thermal entropy in the small $L_{\tau}$ regime, where the weak-coupling expansion is valid. 
We show that the result coincides with that for $N-1$ free complex scalar fields, 
which is in good agreement with the analytical prediction \cite{Monin:2015xwa}
based on the free energy in the large-$N$ limit.

{\bf Model and Lattice setup}:
The continuum bare action of the ${\mathbb C}P^{N-1} = 
{\rm SU}(N)/({\rm SU}(N-1)\times {\rm U}(1))$ sigma models 
(without the topological $\theta$-term) is
$S={1\over{2g^{2}}} \int d^2 x |D_\mu \phi|^{2}\,$
with $|\phi|^2 =1$, $D_\mu \phi= (\p_\mu + iA_\mu) \phi$.
Here, 
$\phi=(\phi^1,\cdots,\phi^N)$ is 
an $N$-component complex scalar field, and
 $A_\mu$ is an auxiliary U($1$) gauge field defined as 
$A_\mu \equiv \frac{i}{2} \bar{\phi} \cdot 
\overset{\leftrightarrow}{\p}_\mu \phi$. 
This model has a PSU($N$)=SU($N$)/${\mathbb Z}_{N}$ 
global symmetry, 
where the ${\mathbb Z}_{N}$ center is removed  
since it coincides with a subgroup of U($1$) gauge symmetry 
and is redundant.

On the lattice, the action is expressed as 
\cite{Berg:1981er,Campostrini:1992ar,Alles:2000sc,Flynn:2015uma,Abe:2018loi}
\begin{equation}
S=N\beta \sum _{n,\mu} \left( 2- \bar{\phi}_{n+\mu}\cdot \phi_{n} \,
\lambda_{n,\mu} - \bar{\phi}_{n}\cdot \phi_{n+\mu} 
\bar{\lambda}_{n,\mu} \right)\,,
\label{eq:latt-action}
\end{equation}
where $\phi_{n}$ is an $N$-component complex scalar field satisfying 
$\bar{\phi relation is one-to-one }_{n} \cdot \phi_{n} =1$ and $\lambda_{n,\mu}$ 
is a link variable corresponding to the auxiliary U($1$) gauge 
field ($\lambda_{n,\mu}=e^{i A_{\mu} (n)}$). 
Here, $n=(n_x, n_\tau)$ labels the sites on the lattice and 
$(n_x, n_\tau)$ run as $n_x = 1, \cdots, N_s$ and 
$n_\tau = 1, \cdots, N_\tau$, respectively. 
We also note that $N\beta$ corresponds to the inverse of the bare coupling ${1\over{g^{2}}}$.
The advantage of this expression is that the fields can be updated locally in Monte Carlo simulation.
We here adopt the over-heat-bath algorithm \cite{Campostrini:1992ar} to update the fields.

The spacetime geometry on the lattice is 
$\mathbb{T}^2= S^1_s \times S^1_\tau$, where $S_s^{1}$ and $S^1_\tau$ 
have the circumferences $L_{s} = N_{s} a$ and $L_{\tau} = N_{\tau} a$, 
respectively.
According to the renormalization group, the following relation 
between the lattice parameter $\beta$ and the lattice spacing $a$ holds
$\Lambda_{\overline{MS}} \,a = (2\pi\beta)^{{2\over{N}}} e^{-2\pi\beta}\,,$
where $\Lambda_{\overline{MS}}$ is defined as a scale at which 
the renormalized coupling in the $\overline{MS}$ scheme diverges. 
The lattice $\Lambda$ scale $\Lambda_{lat}$ depends on the 
explicit form of the lattice action. 
Comparing $\Lambda_{\overline{MS}}$ in Ref.~\cite{Campostrini:1992ar} 
and $\Lambda_{lat}$ for Eq.(\ref{eq:latt-action}), we find
\begin{equation}
\Lambda_{lat} \, a = {1 \over{\sqrt{32}}} (2\pi\beta)^{2\over{N}} 
e^{-2\pi\beta -{\pi\over{2N}}}\,.
\label{eq:beta-a-relation}
\end{equation}
It gives $a$ for a given $\beta$ for each $N$ with $\Lambda_{lat}$ as a reference scale. This relation is valid for $\beta \gtrsim 1/(\pi N)$, which is comfortably satisfied in this work.

We confirm that the action density $\langle E \rangle =\langle 2  -\bar{\phi}_{n+\mu}\cdot \phi_{n} \lambda_{n,\mu} - \bar{\phi}_{n}\cdot \phi_{n+\mu} \bar{\lambda}_{n,\mu} \rangle$ in our numerical calculations  is consistent with the results based on the strong-coupling expansion $\langle E \rangle \approx 2(1-\beta)$ for low $\beta$ ($\beta \lesssim 0.4$), while it agrees with the result based on weak-coupling expansion $\langle E \rangle \approx 1/(2\beta)$ for high $\beta$ ($2.0 \lesssim \beta$).

By setting $L_s \gg L_\tau$, 
we can approximately simulate the model on $\mathbb R \times S^1$, 
where the compactified circumference $L_{\tau}$ is 
interpreted as an inverse temperature $1/T$.
We will mainly use $L_{\tau}$ in this Letter, 
where the smaller $L_{\tau}$ 
(the higher $\beta$ with fixed $N_\tau$) corresponds to the higher $T$.
The lattice size in this work is mainly $(N_{s},N_{\tau})=(200,8)$.
We also vary $N_{s}$ between $40$ and $200$ to look into the finite-volume effects.
It is notable that the $N_{s} \to \infty$ limit corresponds to a thermodynamic limit, where the model is defined on ${\mathbb R} \times S^{1}$ and the genuine phase transition can occur.
We adopt parameters as $N=3,5,10,20$ and $0.1 \leq \beta \leq 3.9$.

{\bf Deconfinement and Polyakov loop}:
The ground state expectation value of the Wilson 
loop $ W({\mathcal C}) = {\mathcal P} \exp(i\oint_{\mathcal C} A) $ 
is expected to exhibit the exponential area law 
and perimeter law for a large rectangle with space $\hat{R}$ and Euclidean time 
$\hat{T}$
\begin{equation}
\langle W({\mathcal C}) \rangle \,=\, 
C e^{-\sigma \hat{R} \hat{T} 
- \rho (\hat{R} + \hat{T}) } \,,
\label{eq:wilson_loop}
\end{equation}
with the Abelian string tension $\sigma\ge 0$, a constant 
$\rho\ge 0$ of the perimeter term, and a constant $C$. 
The confinement of electrically charged particles 
is defined by the nonvanishing $\sigma$. 
Actually, on lattice simulation with a large $N_s = N_\tau$,
the value of the string tension can be calculated by the large Wilson 
loop~\cite{Campostrini:1992ar}.
If we compactify the spacetime as $\tau\sim \tau+L_\tau$ and impose the PBC, 
the Wilson loop becomes a correlator of Polyakov loops,
 $P(x) \equiv {\mathcal P} \exp(i\int_0^{L_\tau} d\tau A_\tau)_{x}$ 
at $x$ ,
\beq
\langle W({\mathcal C}) \rangle \,=\, 
\langle P(\hat{R}) P^\dag (0) \rangle .
\eeq
Since the Wilson loop (\ref{eq:wilson_loop}) satisfies the 
clustering property 
$\langle P(\hat{R}) P^\dag (0) \rangle 
 \to |\langle P \rangle|^2$ in $\hat{R}\rightarrow \infty$, 
the confinement $\sigma\not=0$ necessitates the 
vanishing Polyakov loop $\langle P \rangle=0$. 
The ground state expectation value of the Polyakov loop $\langle P \rangle$ 
is a better observable for the confinement-deconfinement transition 
in the $L_\tau \ll L_s$ system, where taking the large Euclidean time is technically difficult.
This situation is parallel to four-dimensional QCD with fundamental quarks.

On the lattice, the Polyakov loop
is expressed as the product of the link variable,
\begin{equation}
P\equiv {1\over{N_{s}}} \sum_{n_x}\prod_{n_\tau} \lambda_{n,\tau}.
\end{equation} 
\begin{figure}[t]
\centering
 \includegraphics[width=1.0\linewidth]{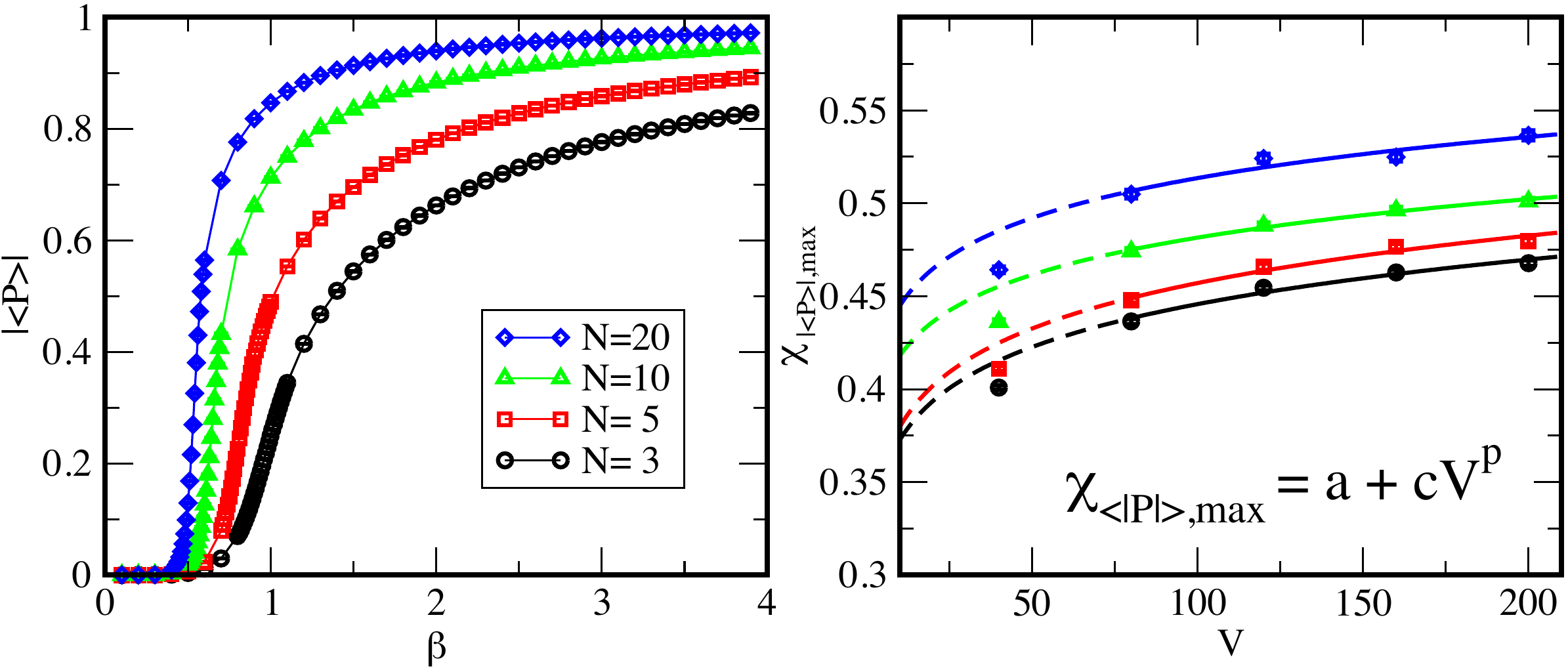}
 \caption{(Left) The absolute value of expectation value of Polyakov loop $|\langle P \rangle|$ as a function of $\beta$ 
(Right) The volume dependences of the maximal peak height of the Polyakov loop susceptibility for each $N$ by varying $N_{s}$ as $N_{s}=40,80,120,160,200$ with $N_{\tau}=8$. 
 }
\label{fig:Poly}
\end{figure}
The results for $|\langle P \rangle|$ as a function of $\beta$ 
for $N=3,5,10,20$ are summarized in the left panel of Fig.~\ref{fig:Poly}.
Here, the lattice parameters are fixed by ($N_s,N_\tau$) $=$ ($200,8$).
It clearly shows $|\langle P \rangle| \approx 0$ for low 
$\beta$ (large $L_{\tau}$) and $|\langle P \rangle| \not= 0$ for 
high $\beta$ (small $L_{\tau}$). For intermediate $\beta$, the 
value of $|\langle P \rangle|$ gradually increases especially 
for small $N$, as is consistent with a crossover behavior.

The corresponding susceptibility of $\langle |P|\rangle$ has a peak, 
and then we define the critical length (or the critical inverse temperature) for each $N$ from the peak position of $\beta$. We also investigate the heat capacity, $C_v = (E -\langle E \rangle)^2  N_\tau^2$, where $E$ denotes the action density. The heat capacity for each $N$ has the peak at the same value of $\beta$ with the one for the susceptibility of $\langle |P|\rangle$.

To see the strength of the transition more clearly, we also investigate the volume dependence of the peak value of the Polyakov loop susceptibility, $\chi_{\langle | P| \rangle}= V (\langle | P|^2 \rangle- \langle | P| \rangle^2)$, by varying $N_{s}$ as $N_{s}=40,80,120,160,200$ with $N_{\tau}=8$ fixed.  
We study the scaling with respect to volume, $V=N_s$.
 We fit the four data points with the large volume, $N_s = 80$--$200$, by a function
$\chi_{\langle|P|\rangle,{\rm max}} \,=\, a + c V^{p}\,$
as shown in the right panel of Fig.~\ref{fig:Poly}.
The best fit values of the exponent are $p= 0.056(7), 0.058(7), 0.052(7),$ and $0.043(8)$ for $N=3,5,10$, and $20$, respectively.
Since it is known that $p=1$ indicates the first-order transition while $0<p<1$ indicates the second-order or crossover transitions \cite{Fukugita:1990vu}, this result supports our argument that the order of the transition is crossover for finite $N$.
Furthermore, all results of the exponent are consistent with each other within $2$--$\sigma$ statistical error, so that we conclude that there is no clear $N$-dependence for the strength of the transition in these finite $N$ analyses.

\begin{figure}[t]
\centering
 \includegraphics[width=0.6\linewidth]{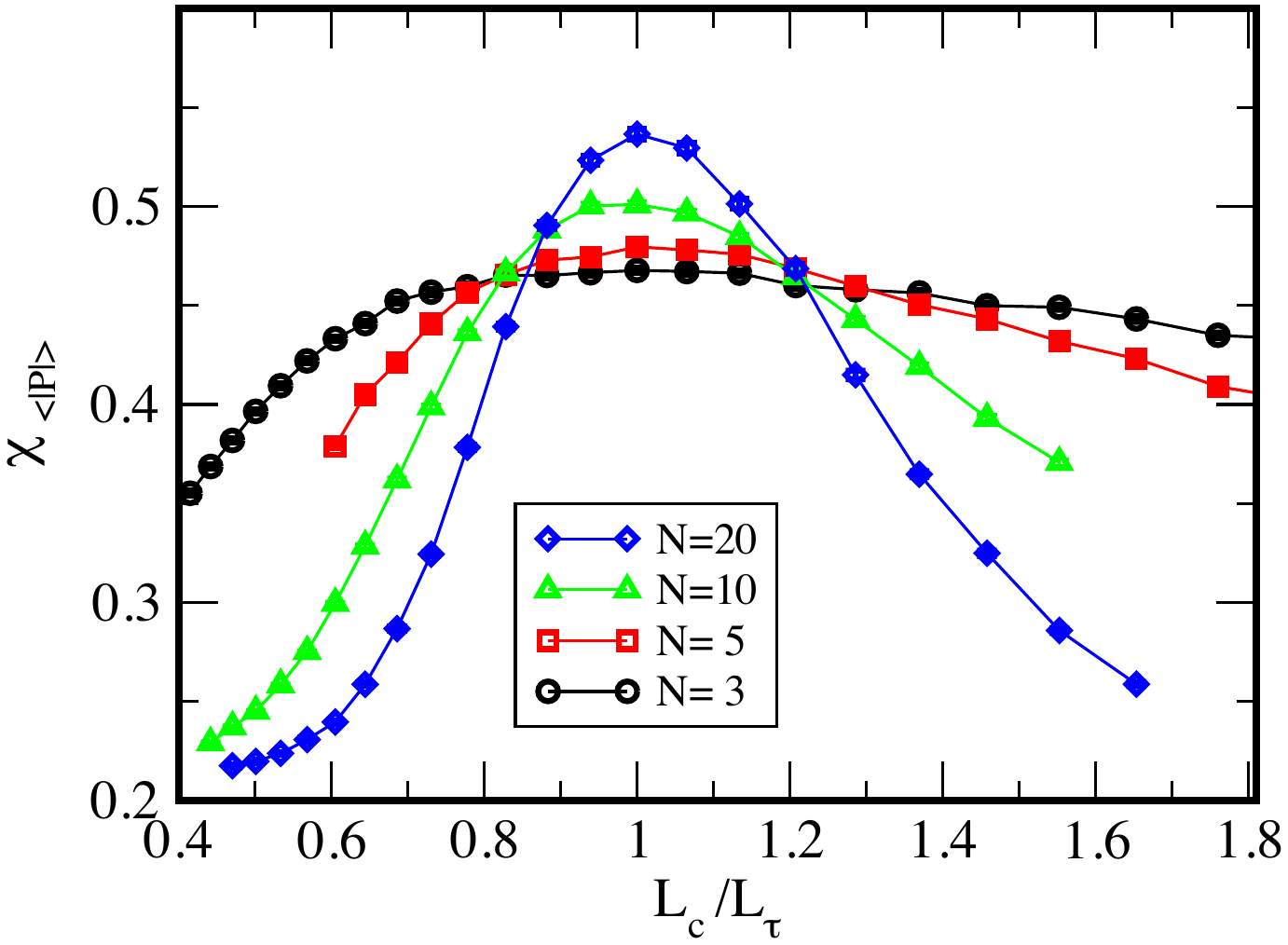}
 \caption{ The susceptibility of the expectation value of absolute value of Polyakov loop $\langle |P|\rangle$ as a function of $L_{c}/L_{\tau}$ for $N=3,5,10,20$ with ($N_s,N_\tau$)$=$($200,8$).
 }
\label{fig:peak}
\end{figure}

On the other hand, in the large-$N$ limit, we first take the large-$N$ limit with a finite $L_\tau$.
To explore the $N$ dependence of the strength of the transition at a finite fixed-volume, the susceptibility of $\langle |P|\rangle$ as a function of a linear scale of $1/L_\tau$  is shown in Fig.~\ref{fig:peak}.
Here, the critical length ($L_c$) for each $N$ is defined from the peak position of $\beta$ with fixed ($N_s,N_\tau$)=($200,8$) simulations, and $\beta$ is translated into the length $L_\tau=N_\tau a$ via Eq.~(\ref{eq:beta-a-relation}). The $N$ dependence of the susceptibility indicates that the peak is quite broad for small $N$ but it gets sharper as $N$ increases. This result suggests that the order of transition is crossover for finite $N$ while it is possibly transformed into a phase transition in the large-$N$ limit as conjectured in Ref.~\cite{Monin:2015xwa}.

{\bf Global PSU($N$) symmetry}:
It was claimed in Ref.~\cite{Monin:2015xwa} that the deconfinement phase transition is associated with the PSU($N$) symmetry breaking in the large-$N$ analysis while, at finite $N$, the PSU($N$) global symmetry  is never broken in two-dimensions even at finite temperature because of the CMW theorem. Now, we found the confinement and deconfinement phases even for finite $N$, then the questions arise: whether the PSU($N$) symmetry exists in the deconfinement phase and, it it exists, how the symmetry is realized in the phase.

\begin{figure}[t]
 \includegraphics[width=1.0\linewidth]{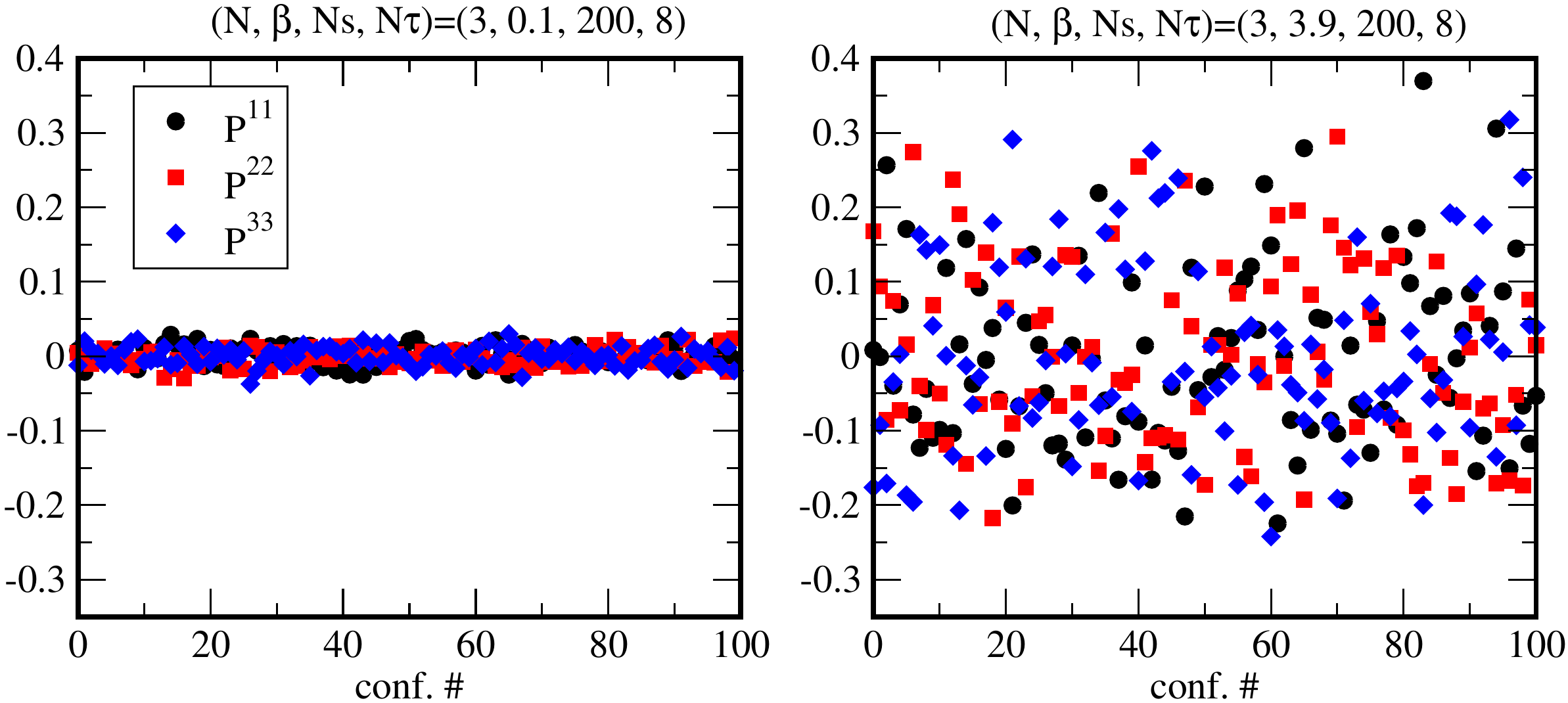}
 \caption{$P^{ii}$ ($i=1,2,3$) with $N=3$ for $\beta=0.1$ 
(confinement) (left) and $\beta=3.9$ (deconfinement) (right) are shown.
The horizontal axis stands for the label number of configurations and we pick up one per $5000$ sweeps.}
\label{fig:Pij}
\end{figure}

To look into this property, we calculate the following 
$N\times N$ matrix quantity,
\begin{equation}
P^{ij} \equiv \sum_{n} \bar{\phi}^{i} \phi^{j} (n) - {1\over{N}} \delta^{ij}\,,~~~
i,j=1,...,N
\end{equation}
whose expectation value serves as an 
order parameter of the PSU($N$) symmetry in the ${\mathbb C}P^{N-1}$ model.
The distributions of the diagonal components $P^{ii}$ ($i=1,2,3$) with 
$N=3$ for the confinement phase ($\beta=0.1$) 
and the deconfinement phase ($\beta=3.9$) are presented in Fig.~\ref{fig:Pij}.
The horizontal axis stands for the label number of configurations, 
where we pick up one configuration per $5000$ sweeps.
In the confinement phase, the values of $|P^{ii}|$ are relatively small 
for all the configurations and lead to $\langle P^{ii}\rangle \approx 0$ 
as $\langle  P^{11}\rangle = -2.80(5808)\times 10^{-5}, \langle  P^{22}\rangle 
= 3.31(553) \times 10^{-4}, \langle  P^{33}\rangle = -3.03 (589) \times 10^{-4}$.
On the other hand, in the deconfinement phase, the values of $P^{ii}$ for some of configurations 
are far from zero and are distributed broadly.
The expectation values for this case is, however, still consistent 
with zero, where $\langle  P^{11}\rangle = -8.22\times 10^{-4}, 
\langle  P^{22}\rangle = 1.48 \times 10^{-6}, \langle  P^{33}\rangle 
= 8.21 \times 10^{-4}$ with $O(10^{-2})$ statistical errors. 
We can phrase that PSU($N$) 
symmetry is realized at a ``quantum level'' 
in the deconfined phase , 
in contrast to the confinement phase in which 
it is realized at a ``classical level.''
The word ``quantum" is to emphasize $P^{ii}$ vanishes only after taking ensemble average while the field variables on each configuration are ordered.
We carry out the similar analyses also for $N=5,10,20$ and find no sign of the PSU($N$) symmetry breaking although the fluctuation seems to get larger with $N$.
It is still an open question whether or not this global symmetry is broken in the large-$N$ limit.

{\bf Thermal entropy density}:
Now, we numerically find all $N$ components of $\phi^i$ are equivalent even in the deconfinement phase, but actual degrees of freedom must be $N-1$ due to one constraint, $|\phi|^2=1$.
To show it manifestly, let us study the thermal entropy density ($s$), which counts the degrees of freedom of the system, in the deconfinement phase.

In the finite temperature (quenched) QCD, the thermal entropy has been calculated by two independent ways; from 
the energy-momentum tensor (EMT) and the free energy.
It has been confirmed that these approaches give consistent results~\cite{Asakawa:2013laa}.
We first focus on the EMT followed by the free energy.
We define the following quantities as a lattice EMT:
\beq
T_{\tau\tau}&=&2 N\beta (2- \bar{\phi}_{n+\tau}\cdot \phi_{n}\lambda_{n,\tau} 
- \bar{\phi}_{n}\cdot \phi_{n+\tau}\bar{\lambda}_{n,\tau}) \nonumber\\
&&-(\mbox{trace part}).
\eeq
 $T_{xx}$ can be defined as well.
The vacuum expectation value of the trace part is subtracted, in a manner parallel to the lattice EMT for the $O(N)$ sigma model~\cite{Makino:2014sta,Makino:2014cxa}.

Here, we use the bare coupling constant instead of calculating the renormalized EMT, since it is a good approximation  in the weak coupling regime.
The thermal entropy density is given by $T_{xx}-T_{\tau\tau} =  sT$ 
with $T\equiv 1/L_{\tau}$ in the thermodynamic limit, 
where the divergent part of the EMT is cancelled between the two terms.

\begin{figure}[t]
\centering
 \includegraphics[width=0.6\linewidth]{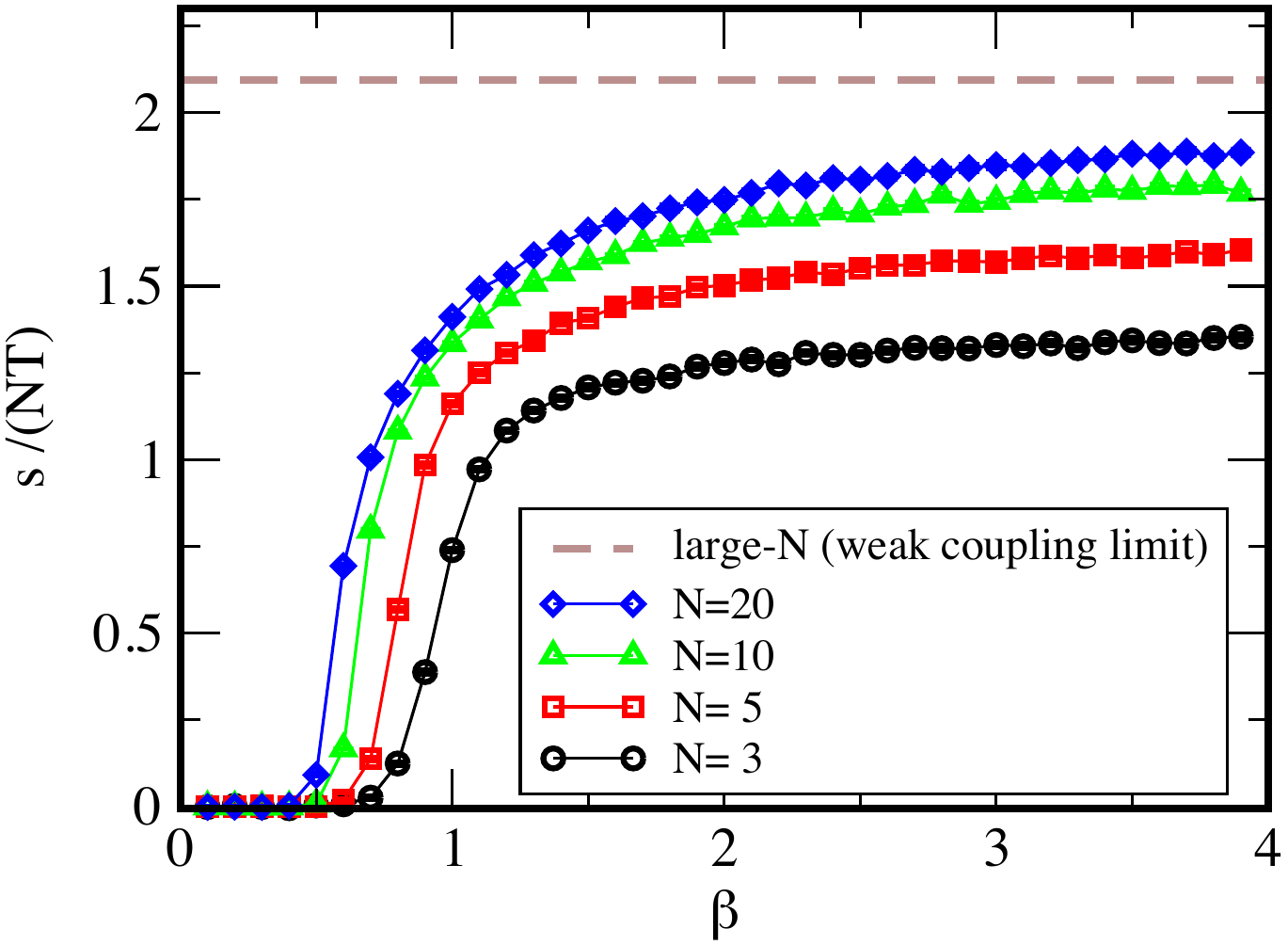}
 \caption{Thermal entropy density for single $\phi$ ($s/(NT)= N_{\tau}^{2}\langle T_{xx}-T_{\tau\tau}\rangle/N$) for $N=3,5,10,20$. The dotted line denotes the large-$N$ results in small $L_\tau$ regime, $2\pi /3$.}
\label{fig:entropy}
\end{figure}

The results of the thermal entropy density for single scalar field
for $N=3,5,10,20$ as a function of $\beta$ are shown in Fig.~\ref{fig:entropy}.
The thermal entropy density becomes non-zero around a certain $\beta$ corresponding to $L_c$ and monotonically grows up in the deconfinement phase. 
For high-$\beta$ regions, the $\beta$ dependence gets gentler for 
each $N$, where we fit them by a function $g(\beta) = a + b/\beta$ 
between $3.0 \le \beta \le 3.9$. 
The best fit values of $a$ are $a_{N=3}=1.418(27), a_{N=5}=1.681(26), a_{N=10} =1.889(29), 
a_{N=20}= 2.024(30)$.
We then find that the values in the $\beta \rightarrow \infty$ limit are consistent with $2\pi(N-1)/(3N)$.

On the other hand, the free energy density for a free massive complex scalar field in the finite temperature ($T=1/L_\tau$) is given by 
\beq
f = \frac{1}{L_s L_\tau} \sum_{n=-\infty}^\infty \log 4 \sinh^2 \frac{\omega_n L_\tau}{2}   -f_0
\eeq
from the analytical calculation (see Appendix.~A). Here, $\omega_n^2 = \left( \frac{2\pi n}{L_s} \right)^2 + m^2$ and $f_0$ denotes the counter term which cancels the UV divergence.
Then, the thermal entropy density in the massless and thermodynamic limit ($L_s \rightarrow \infty$) is given by $s/T = - \frac{1}{T} \frac{\partial f}{\partial T} = \frac{2\pi}{3}$ for a single complex scalar field.
Our numerical results indicate that the actual degree of freedom of 
the $\mathbb{C}P^{N-1}$ model is ($N-1$) massless free complex scalar fields in the deconfinement phase.
Furthermore, the large-$N$ limit of our results is consistent with the prediction calculated from the free energy for the large-$N$ limit in small $L_{\tau}$ regime, $f= -\frac{N \pi}{3 L^2_\tau}$~\cite{Monin:2015xwa,Luscher:2004ib,Teper:2009uf,Caselle:2015tza} using similar calculations.

{\bf Summary and Discussion}:
In this Letter, we have reported the results on the non-perturbative 
aspects of the ${\mathbb C}P^{N-1}$ model on $S^{1}({\rm large}) 
\times S^{1}({\rm small})$: 
We have found a confinement-deconfinement 
crossover by calculating the $L_{\tau}$ dependence of the expectation 
value of the Polyakov-loop, where the peaks of its susceptibility 
get shaper as $N$ increases.
We have clearly shown that the global PSU($N$)=SU($N$)$/{\mathbb Z}_{N}$
symmetry remains unbroken at ``quantum" and ``classical" levels  
for the small and large $L_{\tau}$, respectively,
consistent with the CMW theorem.
We have obtained the thermal entropy in small $L_{\tau}$ regime for small and 
large $N$, and have shown that the large $N$ values agree with 
the small $L_\tau$ results of the large-$N$ approximation.

Our results give a new insight on the phase diagram of the $\C P^{N-1}$ model. 
Furthermore, since some of the conjectures we have discussed originate in four-dimensional gauge theories, 
our results also would give significant implications to four-dimensional gauge theories.

As a future avenue, our formalism can be extended to 
the model with different geometries and/or boundary conditions, 
such as the model on $\R \times S^{1}$ with $\Z_{N}$ twisted boundary conditions, where $\Z_{N}$ symmetry 
is exact \cite{Dunne:2012ae,Dunne:2012zk,Tanizaki:2017qhf}, 
and the model on a finite interval for which the Casimir effect is extensively argued  
 \cite{Milekhin:2012ca,Bolognesi:2016zjp,Milekhin:2016fai,Betti:2017zcm,Flachi:2017xat,Bolognesi:2018njt,Chernodub:2019nct,Pavshinkin:2019bed,Flachi:2019jus}.
For the former, whether it undergoes a first-order phase transition or has adiabatic continuity of the vacuum structure \cite{Sulejmanpasic:2016llc} and whether fractional instantons have physical consequences \cite{Eto:2004rz,Eto:2006mz,Eto:2006pg,Bruckmann:2007zh,Brendel:2009mp,Nitta:2014vpa, Nitta:2015tua, Bruckmann:2018rra, Itou:2018wkm,Misumi:2014jua,Misumi:2015dua,Wan:2018zql} in the model are questions attracting a great deal of attention in terms of the resurgence theory of the models \cite{Dunne:2012ae,Dunne:2012zk,Misumi:2014jua,Misumi:2015dua,Buividovich:2015oju,Fujimori:2016ljw,Fujimori:2017oab,Fujimori:2017osz,Dorigoni:2017smz,Fujimori:2018kqp}.


\begin{acknowledgements}
{\bf Acknowledgements:} This work is supported by the Ministry of Education, Culture, 
Sports, Science, and Technology(MEXT)-Supported Program for the 
Strategic Research Foundation at Private Universities ``Topological 
Science" (Grant No. S1511006) 
and 
by the Japan Society for the Promotion of Science (JSPS) 
Grant-in-Aid for Scientific Research (KAKENHI) Grant Number 
(18H01217).
This work is also supported in part by JSPS KAKENHI Grant Numbers 
19K03875 (E.\ I.), 18K03627 (T.\ F.), 19K03817 (T.\ M.), and  16H03984 (M.\ N.).
The work of M.\ N. is also supported in part 
by a Grant-in-Aid for Scientific Research on Innovative Areas 
``Topological Materials Science" (KAKENHI Grant No. 15H05855) 
from MEXT of Japan. 
Numerical simulations were performed on SX-ACE at the Research Center for Nuclear Physics (RCNP), Osaka University and TSC at Hiyoshi department of Physics, Keio University.
\end{acknowledgements}

\section{Appendix A: Thermal entropy of a free massive scalar field}

In this appendix, we calculate the thermal entropy of 
a free massive scalar field
\beq
S = \int d^2 x \, (|\p_\mu \phi|^2 + m^2 |\phi|^2). 
\eeq
On a torus with periods $(L_\tau, L_s)$, 
this model can be regarded as 
a collection of infinitely many 2D harmonic oscillators 
with frequencies $\omega_n^2 = \left( \frac{2\pi n}{L_s} \right)^2 + m^2$ at temperature $T = 1/L_\tau$,
so that the partition function is given by 
\beq
Z = \prod_{n = - \infty}^{\infty} \frac{1}{4\sinh^2 \frac{L_{\tau} \omega_n}{2}}.
\eeq
The free energy density can be obtained from $Z=e^{-L_\tau L_s f}$ as
\beq
f = \frac{1}{L_\tau L_s } \sum_{n = - \infty}^{\infty} \log 4\sinh^2 \frac{L_{\tau} \omega_n}{2} + m^2 - f_0,
\eeq
where the last term denotes the counter term 
which cancels the UV divergence.
In the infinite volume limit $L_s \rightarrow \infty$, 
the summation over the Kaluza-Klein momentum 
is replaced by the momentum integration
\beq
f = \frac{1}{L_\tau} \int \frac{dk}{2\pi} \log 4\sinh^2 \frac{L_\tau \sqrt{k^2 + m^2}}{2} - f_0. 
\eeq
The energy density can be calculated from this free energy as
\beq
\epsilon \, = \, \frac{\p}{\p L_\tau} (L_\tau f) = \int \frac{dk}{2\pi}  \sqrt{k^2+m^2} \,\coth{\frac{L_\tau}{2} \sqrt{k^2 + m^2}} - f_0, \nonumber\\
\eeq
From these expression, 
we find that the high temperature (small $L_\tau$) behavior of 
the thermal entropy density in the infinite volume limit takes the form
\beq
s ~=~ L_\tau(\epsilon-f) ~=~\frac{1}{L_\tau} \left[ \frac{2\pi}{3} + \mathcal O(L_\tau m) \right]. 
\eeq
It is notable that $s$ is independent of the choice of the counter term. 
Since the pressure in the infinite volume limit can be written as
\beq
P ~=~ - \frac{\p}{\p L_s} (L_s f) ~=~ - f,
\eeq
the thermal entropy density can also be written as $s = L_\tau (\epsilon + P)$.


\end{document}